\documentclass[%
 reprint,
nofootinbib,
 amsmath,amssymb,
 prl,
]{revtex4-2}
\usepackage{graphicx}% Include figure files
\usepackage{dcolumn}% Align table columns on decimal point
\usepackage{bm}% bold math
\usepackage{hyperref}% add hypertext capabilities

\usepackage{aas_macros}

\begin{document}

\preprint{APS/123-QED}

\title{Kinematic Lensing Ratio: \texorpdfstring{\\}{ }Reviving Weak Lensing Cosmography as a Geometric Dark Energy Probe}

\author{Qinxun Li}\email{qinxun.li@utah.edu}
\affiliation{%
  Department of Physics and Astronomy, The University of Utah, 115 South 1400 East, Salt Lake City, UT 84112, USA
}%

\date{\today}

\begin{abstract}
  We introduce the kinematic lensing ratio (KiLeR), a geometric dark-energy probe from weak lensing. Combining shear ratios with intrinsic galaxy shapes inferred from kinematics, KiLeR naturally mitigates most first-order lensing systematics, including redshift errors, intrinsic alignments, and baryonic effects. The forecast on \textit{Roman} shows a $192\%$ improvement in dark energy constraints from adding KiLeR, providing independent tests of dark energy evolution. We quantify the requirements on systematic and statistical errors and discuss the pathways towards KiLeR observation.
\end{abstract}

\maketitle

\section{Weak Lensing Cosmography}

The nature of cosmic acceleration remains one of the most profound mysteries in fundamental physics (for a review, see \cite{Weinberg:2013agg}).
While the standard $\Lambda$CDM model, which interprets the acceleration as driven by Einstein's cosmological constant ($\Lambda$), has been remarkably successful in the era of the Planck satellite \cite{Planck-2018-overview,Planck-2018-cosmology}, recent Dark Energy Spectroscopic Instrument (DESI) analysis \cite{DESI.DR2.BAO.lya,DESI.DR2.BAO.cosmo,2025PhRvD.112h3529G} reported $\sim 3-4\sigma$ evidence of evolving dark energy. This potential discovery marks a pivotal shift in community focus from constraining the density parameter $\Omega_{\rm DE}$ to characterizing the time-evolution of the equation of state, $w(z)$.

As Baryon Acoustic Oscillation (BAO) measurements at low redshifts ($z<1$) approach the cosmic variance uncertainty floor, independent and robust geometric probes become critical. Such probes are crucial for validating the DESI finding, which shows strong dependence on the calibration of Type Ia supernovae (SN) \cite{DES2025DovekieCosmology}.
Moreover, the DESI hint can also be interpreted as a negative neutrino mass \cite{2024JHEP...09..097C,Elbers2025}, decaying dark matter \citep[e.g.][]{Lynch-Knox2025,2025JCAP...07..059C}, non-zero spatial curvature \cite{2025JCAP...08..014C}
or early dark energy \cite{2025PhRvD.112f3548C}.
Tight constraints on the expansion history at the matter-dominated era are then desired to break degeneracies between these models and dark energy \citep[e.g.][]{2025PhRvD.112h3529G,DESI2024.VI}. Weak gravitational lensing (WL) measures both the expansion history and the structure growth through the distortion and magnification of background sources \citep[for reviews, see][]{Kilbinger2015,Mandelbaum2018}. The tangential shears $\gamma_t$ of background galaxies are determined by the geometric lensing efficiency and the lens mass distribution $\Sigma(\theta)$,
\begin{equation}
  \gamma_t(\theta) \propto
  \underbrace{\frac{D_{ls}D_{l}}{D_{s}}}_{\text{Geometry}}
  \times
  \underbrace{\left[\langle\Sigma(<\theta)\rangle-\Sigma(\theta)\right]}_{\text{Mass distribution}}
\end{equation}
where $D_{l}$, $D_{s}$, and $D_{ls}$ are the angular diameter distances to the lens, to the source, and between them, respectively. However, the potential of WL to constrain dark energy is far from fully realized \footnote{Although the DES Y3 3$\times$2 pts has shown an ability to replace the Cosmic Microwave Background (CMB) data in constraining the evolution of dark energy \cite{DESI.DR2.BAO.cosmo}, the improvement becomes mild when BAO is already combined with CMB. } due to our limited understanding of WL systematics and the non-linear structure formation with baryonic feedback \citep[e.g.][]{DESY3WLw0wa,2024arXiv241022272M,2025arXiv251204209B}.
Therefore, decoupling the geometric and the structure growth information provides more constraining power and even more insights into the nature of dark energy.

Weak lensing cosmography, also known as the Shear Ratio (SR) test, has long been proposed as a promising DE probe \cite{Jain2003,Bernstein2004,Taylor2007}. By taking ratios of the lensing shears at different source redshifts behind the same lens, the dependence on the lens mass distribution effectively cancels out, yielding a purely geometric observable,
\begin{equation}
  R(z_l;z_{s1},z_{s2})\equiv \frac{\gamma_t(\theta|z_l,z_{s1})}{\gamma_t(\theta|z_l,z_{s2})}=\frac{D_{ls1}D_{s2}}{D_{ls2}D_{s1}}.
\end{equation}
This decoupling renders SR robust against small-scale baryonic effects and uncertainties in matter clustering, theoretically enabling the use of high signal-to-noise (SNR) measurements on small scales.

Despite this theoretical elegance, SR is remarkably sensitive to photometric redshift (photo-z) errors and intrinsic alignment (IA) systematics \cite{Kitching2008}. These drawbacks have historically stifled its utility as a primary dark energy probe. In recent Stage-III WL surveys, SR is typically relegated to a complementary role, used merely to constrain nuisance parameters associated with photo-z and IA (e.g. \cite{Sanchez2022, Emas2024, Emas2025, Rana2025}).

In this work, we introduce the Kinematic Lensing Ratio (KiLeR) method and demonstrate that, by incorporating kinematic information \cite{Blain2002,Morales2006}, KiLeR resolves the dominant systematics of SR and revives its potential as a robust and powerful dark energy probe, effectively re-establishing weak lensing cosmography in the era of Stage-IV surveys.

\section{Kinematic Lensing Ratio}

Standard WL relies on the statistical assumption of random galaxy orientations to overcome the degeneracy between the lensing shear $\gamma$ and the intrinsic galaxy shape $e_{\rm int}$. The observed ellipticity is $e_{\rm obs} \approx e_{\rm int} + \gamma$, where the dispersion of $e_{\rm int}$ (shape noise) necessitates averaging over large ensembles of galaxies ($N\propto 1/\sigma^2$ for target precision $\sigma$) for meaningful shear measurements.

\textbf{Kinematic Lensing (KL)} breaks this degeneracy by inferring the intrinsic shape $e_{\rm int}$ individually from the rotation of disk galaxies \cite{2002ApJ...570L..51B, 2006ApJ...650L..21M,Huff2013,2015MNRAS.451.2161D}.
For the shear component aligned with the galaxy major axis ($\gamma_+$), KL utilizes the Tully-Fisher Relation \citep[TFR;][]{TullyFisher} to constrain the disk orientation and the intrinsic shape.
The TFR predicts the intrinsic circular velocity $v_{\rm TF}$ from the galaxy's absolute magnitude $M_B$
\begin{equation}
  \log_{10} v_{\rm TF} = a + b (M_{\rm B}-M_{\rm p}).
\end{equation}
By comparing $v_{\rm TF}$ with the observed rotation velocity $v_{\rm obs}$, we derive the inclination angle $\sin i = v_{\rm obs}/v_{\rm TF}$.
Combined with a prior on the disk thickness $q_z$, the unlensed intrinsic ellipticity is reconstructed as
\begin{equation}
  |e_{\rm int}| = \frac{(1-q_{\rm z}^2)\sin^2 i}{2-(1-q_{\rm z}^2)\sin^2 i}.
\end{equation}
The shear $\gamma_+$ is then extracted by differencing this kinematically-inferred intrinsic shape from the photometrically observed shape $e_{\rm obs}$
\begin{equation}
  \gamma_+ = \frac{e_{\rm obs}-e_{\rm int}}{2(1-e_{\rm int}^2)}.
\end{equation}

The cross component of the shear ($\gamma_{\times}$, defined as the shear measured $45^\circ$ from the major axis), on the other hand, is measured through the distortion of the velocity field. Lensing induces a misalignment between the photometric and kinematic axes, generating an anomalous gradient velocity $v'_{\rm minor}$ along the lensed photometric minor axis. Leveraging this effect, $\gamma_{\times}$ is inferred as \cite{Huff2013}
\begin{equation}
  \gamma_{\times} = -\frac{v_{\rm minor}'}{v_{\rm TF}}\sqrt{\frac{(1-q_z^2)e_{\rm int}}{2(1+e_{\rm int})}}.
\end{equation}

While KL significantly enhances per-object sensitivity (shape noise) by an order of magnitude \cite{Xu2023,Pranjal2023}, its application to standard 3$\times$2 pts or cosmic shear is still limited by baryonic physics uncertainties in structure growth.
By applying KL to the Shear Ratio test, we propose the \textbf{Kinematic Lensing Ratio (KiLeR)}. KiLeR resolves most problems and inherits the major advantages of both KL and SR, making it a promising dark energy probe:

\begin{itemize}
  \item \textbf{Photo-$z$ Free:} The order-of-magnitude lower shape noise of KL allows us to utilize smaller samples with high-quality spectroscopic redshifts for lensing measurement. This completely eliminates the photometric redshift errors that plague traditional SR.
  \item \textbf{IA Free:} Traditional WL is biased by IA because tidal forces correlate intrinsic orientations, violating the random-phase assumption. KiLeR measures $\gamma$ for each object individually, rendering the measurement independent of the intrinsic correlation of galaxy orientations.\footnote{Although they can leak into the ratio measurement in some secondary way through the TFR calibration, these effects are expected to be second-order. Moreover, the IA of disk galaxies is intrinsically weak.}
  \item \textbf{Baryon Insensitive:} Inheriting the property of the Shear Ratio, the KiLeR observable depends only on the cosmological background \footnote{As we discuss in the End Matter, the magnification effect introduces some mass-distribution dependence back in the small-scale modeling. But we have multiple tools to deal with this.}, cancelling out the sensitivity to the mass profile and baryonic feedback of the lens.
  \item \textbf{Geometric Probe:} KiLeR decouples the expansion history information from structure growth, providing a clearer picture for dark energy model building. It also enables efficient tests for a wider range of models because the calculation of the distance-redshift relation is much faster than that of structure formation, including both theoretical derivations and numerical computations.
\end{itemize}

Realizing KiLeR requires a large sample of high-resolution imaging and spatially resolved galaxy spectra. Although this sounds challenging, such data are within reach of the upcoming Nancy Grace Roman Space Telescope (Roman; \cite{Roman2019}), by extracting kinematics from forward-modeling the $R\sim600$ grism slitless spectra as demonstrated by \cite{Xu2023}. We therefore forecast the KiLeR constraint from Roman surveys.

\section{Forecast}

We forecast the dark energy constraints from the Roman High Latitude Wide Area Survey \citep[HLWAS-Medium;][]{2022ApJ...928....1W,RomanTAC2025} combined with DESI-like spectroscopic lens samples.
Following \cite{Xu2023}, we assume a Roman KL source sample with a shape noise of $\sigma_\gamma = 0.035$ and a number density of $n_g = 4$ arcmin$^{-2}$ distributed over 10 tomographic bins ($0.6<z<3$). These parameters are estimated by applying quality cuts to the images and emission lines of the well-resolved disk galaxies on a deep-field mock catalog, as detailed in the end matter.

\begin{figure}[tb]
  \centering
  \includegraphics[width=246pt]{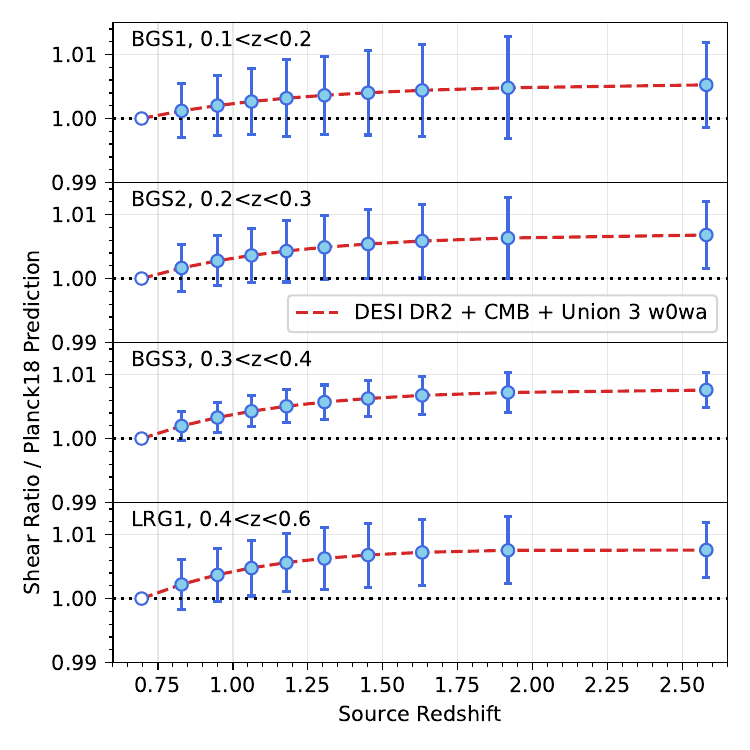}
  \caption{\label{fig:signal} The forecasted KiLeR measurements on the DESI BGS+LRG lens sample with Roman KL sources. We use the first tomographic source bin as the reference bin and show the ratios with respect to it. From top to bottom, the lens redshift bins are $z_l\in[0.1,0.2]$, $[0.2,0.3]$, $[0.3,0.4]$, and $[0.4,0.6]$, respectively. We normalize the ratios to the Planck18 $\Lambda$CDM prediction and plot the input cosmology as the red dashed line for comparison. }
\end{figure}

For the lenses, we adopt the redshift distributions and Halo Occupation Distribution (HOD) parameters of the DESI BGS ($0.1<z<0.4$) and LRG ($0.4<z<0.6$) samples \cite{2024MNRAS.533..589Y,2024MNRAS.530..947Y}, assuming full overlap with the $2,400$ deg$^{2}$ Roman HLWAS-Medium.
Data vectors and analytical covariances are generated using \texttt{AUM} \cite{AUM} and \texttt{OneCovariance} \cite{OneCovariance}, respectively. We maximize the SNR by inverse-variance weighting the ratios measured in 15 log-uniform radial bins ($\theta \in [1, 300]$ arcmin).
Our fiducial cosmology is set to the best-fit $w_0w_a$CDM model from DESI DR2 + CMB + Union3 \cite{DESI.DR2.BAO.cosmo} ($\Omega_{\rm m}=0.3275$, $w_0=-0.667$, $w_a=-1.09$), representing a scenario with evolving dark energy.
As illustrated in Fig.~\ref{fig:signal}, KiLeR measurements achieve sub-percent precision, allowing for a decisive distinction between the input dynamic dark energy model and the Planck18 $\Lambda$CDM prediction.

We perform Markov Chain Monte Carlo forecasts using \texttt{Cobaya} \citep[Code for BAYesian Analysis;][]{Cobaya}, adopting the standard $w_0$-$w_a$ parameterization \cite{CP2001,Linder2003} with priors matching the DESI analysis \cite{DESI2024.VI}, which includes priors of $\omega_{\rm cdm}=\Omega_{\rm cdm} h^2\in \mathcal{U}[0.001,0.99]$, $w_0\in \mathcal{U}[-3,1]$, and $w_a\in \mathcal{U}[-3,2]$.
We combine KiLeR with current state-of-the-art datasets: DESI DR2 BAO \cite{DESI.DR2.BAO.lya,DESI.DR2.BAO.cosmo}, Union3 SN Ia \cite{union3}, and the CMB primary anisotropy from Planck PR4 CamSpec high-$\ell$ TTTEEE \cite{Planck-2020-NPIPE,planck-pr4-comspec} and Planck18 low-$\ell$ TT\&EE \cite{Planck-2018-likelihoods}.
We run 8 chains for each data combination until Gelman-Rubin $R-1<0.01$ convergence \cite{Gelman-Rubin1992}.

\begin{table}[tb]
  \caption{\label{tab:fom} The forecasted $w_0$, $w_a$ uncertainties and the dark energy FoM from KiLeR alone and in combination with other probes.}
  \begin{ruledtabular}
    \begin{tabular}{lccc}
      \textrm{Datasets}   &
      \textrm{$w_0$}  &
      \textrm{$w_a$}  &
      \textrm{FoM}                                                                  \\
      \colrule
      KiLeR & $-0.636_{-0.131}^{+0.155}$ & $-1.38_{-0.90}^{+0.85}$ & 34.6 \\
      KiLeR+SN & $-0.698_{-0.070}^{+0.080}$ & $-0.92_{-0.45}^{+0.42}$ & 104 \\
      KiLeR+CMB & $-0.668_{-0.070}^{+0.071}$ & $-1.07_{-0.33}^{+0.34}$ & 164 \\
      KiLeR+BAO & $-0.663_{-0.076}^{+0.085}$ & $-1.10_{-0.42}^{+0.40}$ & 123 \\
      KiLeR+CMB+SN & $-0.670_{-0.063}^{+0.064}$ & $-1.08_{-0.30}^{+0.30}$ & 198 \\
      KiLeR+CMB+BAO & $-0.668_{-0.050}^{+0.051}$ & $-1.05_{-0.23}^{+0.21}$ & 272 \\
      KiLeR+BAO+SN & $-0.680_{-0.068}^{+0.073}$ & $-1.02_{-0.38}^{+0.36}$ & 153 \\
      KiLeR+CMB+BAO+SN & $-0.675_{-0.047}^{+0.049}$ & $-1.03_{-0.22}^{+0.21}$ & 298 \\
      \hline
      CMB+BAO & $-0.431_{-0.223}^{+0.219}$ & $-1.70_{-0.64}^{+0.63}$ & 37.2 \\
      CMB+BAO+SN & $-0.673_{-0.089}^{+0.092}$ & $-1.05_{-0.31}^{+0.29}$ & 102 \\    \end{tabular}
  \end{ruledtabular}
\end{table}

\begin{figure}[tb]
  \centering
  \includegraphics[width=246pt]{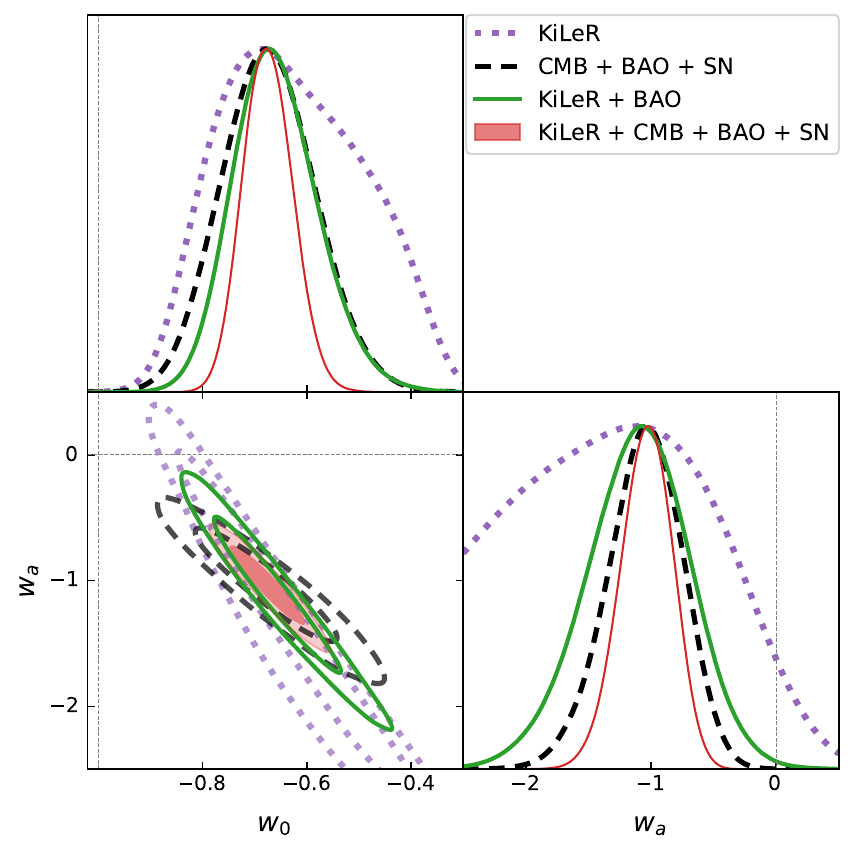}
  \caption{\label{fig:joint} The forecasted dark energy constraints from KiLeR, KiLeR+BAO, and KiLeR+CMB+BAO+SN, shown in purple dotted, green solid, and red filled contours, respectively. The black dashed contours are the state-of-the-art CMB+BAO+SN constraints for comparison. }
\end{figure}

Table~\ref{tab:fom} and Fig.~\ref{fig:joint} present the forecasted constraints. Following the Dark Energy Task Force \cite{Albrecht2006}, we quantify the dark energy constraining power using the Figure-of-Merit (FoM), defined as the inverse area of the $1\sigma$ confidence region in the $w_0$-$w_a$ plane,
\begin{equation}
  \text{FoM} = \frac{1}{\sqrt{\det \text{Cov}(w_0,w_a)}}.
\end{equation}
KiLeR alone yields a dark energy FoM of 34.6, comparable to Roman WL/KL cosmic shear and 3$\times$2 pts \cite{Eifler2021,Xu2023}. We also check the FoM of single-lens-bin KiLeR and find that BGS3 and LRG1 bins contribute the most, with FoM$\sim$8 and 20, while other bins at lower redshifts contribute FoM$<$1. The varying constraining power among lens bins arises from both the distance ratios and the dark energy sensitivity of the distance-redshift relation, promising an even higher FoM from a higher-redshift lens sample. Adding the LRG2 ($0.6<z<0.8$) bin boosts the FoM to $45$ ($201$) for KiLeR (+BAO).

The true power of KiLeR is unlocked in combination with other probes.
When combined with BAO, the FoM leaps to 123, surpassing the current combined power of CMB+BAO+SN (FoM$\sim$102) by $21\%$. The physical origin of this constraining power lies in KiLeR's high-redshift reach. BAO anchors the expansion history at $z<1$ with high precision, providing precise distances to the lenses ($D_l$) and low-$z$ sources ($D_{s1}$). KiLeR then leverages these anchors to tightly constrain the distance-redshift relation of the high-redshift sources $D_{s2}$. This high-$z$ expansion history measurement effectively breaks the degeneracy between the time-varying equation of state $w(z)$ and the dark energy density $\Omega_{\rm DE}$ in the BAO constraint, resolving a primary bottleneck in current observations. Substituting KiLeR for any single probe in the CMB+BAO+SN combination yields $50\%$ to $167\%$ improvements, enabling robust cross-checks against systematics in each individual probe and potentially new physics affecting specific observables.
The full combination of KiLeR+CMB+BAO+SN achieves a remarkable FoM of 298, representing a $192\%$ improvement. We note that this forecast is far from the ultimate limit, as we could go to higher number densities and higher redshifts for the lens sample with Roman data itself.

\section{Impact of Uncertainties}
\label{sec:error}
Although we have tried to take conservative assumptions for the forecast, the large family of unaccounted systematic effects may still degrade the constraining power of KiLeR or bias the inference. For example, the shape noise of KL is estimated under the assumption of perfectly regularized disk morphology and kinematics, whereas the diversity of galaxy kinematic structures can introduce additional noise and bias. The evolution of disk thickness and TFR parameters with redshift \citep[e.g.][]{2008A&A...484..173P,2017ApJ...842..121U,2024A&A...689A.318S} can introduce bias even after calibration. On the observational side, forward modeling of slitless spectroscopy is complicated, as degeneracies among shear, inclination, intrinsic morphology, and kinematic structure can arise and introduce projection bias in the shear, especially at low SNR. While these observational and astrophysical systematics remain to be fully explored, their effects on shear estimates can be parameterized by a multiplicative bias $m$ and an additive bias $c$ following the standard weak-lensing calibration formalism \cite{Mandelbaum2018}:
\begin{equation}
  \gamma_{\rm obs} = (1+m)\gamma_{\rm true} + c + \mathcal{N}(0, \sigma_\gamma^2) + \mathcal{O}(\gamma_{\rm true}^2),
\end{equation}
where the $\sigma_\gamma$ term represents the shape noise. Global additive bias ($c$) is easily removed by enforcing a zero mean shear in each tomographic bin, assuming cosmic isotropy. Multiplicative bias ($m$), however, typically requires rigorous calibration to prevent biased cosmological inference.

\begin{figure}[tb]
  \centering
  \includegraphics[width=246pt]{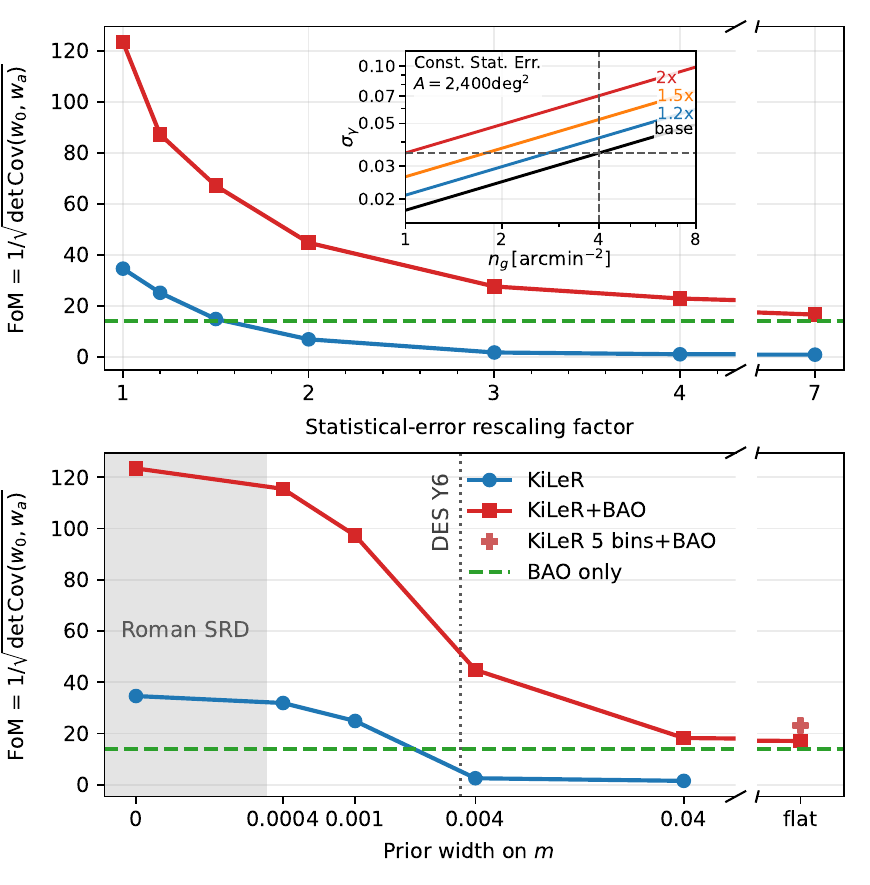}
  \caption{\label{fig:scaling} The forecasted dark energy constraints from KiLeR (blue) and KiLeR+BAO (red) under different assumptions of the statistical error level (top panel) and the width of the Gaussian prior on the multiplicative bias $m$ (bottom panel). The sub-panel in the top panel shows the shape noise $\sigma_\gamma$ and the number density $n_g$ corresponding to the statistical error levels. In the bottom panel, KiLeR+BAO with an extra LRG2 lens bin is plotted in red plus signs to illustrate the self-calibration potential of KiLeR. The $m$ constraint from DES Y6 and the requirement for Roman WL are shown as the vertical dashed and dotted lines, respectively, for comparison. }
\end{figure}

KiLeR naturally self-calibrates $m$ \cite{2016A&A...592L...6S}. A constant $m$ across all source bins cancels out in the KiLeR observable, leaving only the component depending on the redshift of the source bin. Even for this residual, the contamination is disentangled from the cosmological signal because of their different redshift dependencies.
The contamination from $m(z)$ does not depend on the lens redshift, while the cosmological signal has a dramatic lens-redshift dependence, where the dynamical range of KiLeR increases from $\sim 0.1$ to $> 2$ from the lowest to the highest lens redshift bin. This allows KiLeR to self-calibrate $m(z)$ and constrain the cosmological parameters without bias, even under a strong evolution of $m$ with redshift. The self-calibration feature marks KiLeR as a promising probe for next-generation cosmological surveys, especially given the saturation of BAO due to cosmic variance and the increasing challenges of the SN and WL systematics.
However, the self-calibration requires a high SNR. In our baseline, KiLeR alone can only constrain $\Delta m$ to $0.5\%$. Without any calibration, an extra LRG2 lens bin is needed for KiLeR+BAO to outperform BAO alone. Therefore, calibration of $m$ is still necessary for a Roman-like survey. To demonstrate the requirements for calibration, we test a series of Gaussian priors and a wide flat prior on $m$ and find that a $\Delta m \sim 10^{-3}$ level of calibration, which is much weaker than Roman WL \cite{Spergel2015}, is required to prevent significant FoM degradation, as shown in Fig.~\ref{fig:scaling}.

The systematic effects can also enlarge the shape noise or reduce the number density of usable sources, which propagates into a larger statistical error and a weaker FoM. The statistical error, when the shape noise dominates, scales as $\sigma_\gamma/\sqrt{n_{\rm g}A}$, where $A$ is the survey area. We rescale the covariance matrix of the baseline forecast to find the requirements for a Roman-like survey. We find that with $1.5\times$ the baseline statistical error (i.e. $\sigma_\gamma = 0.053$ or $n_{\rm g}=1.8$ arcmin$^{-2}$), the FoM of KiLeR alone drops to BAO level. The $n_{\rm g}$-$\sigma_\gamma$ requirements are shown in the sub-panel of Fig.~\ref{fig:scaling}.
The spatial correlation of shape noise, which acts as IA in WL, may survive the self-calibration and bias the inference. However, these effects are scale-dependent and can be safely mitigated by scale cuts motivated by a scale-independence criterion.

\section{Towards KiLeR cosmology}

By quantifying the uncertainty control requirements, we have formulated the application of KiLeR to observational data as an engineering problem.
The next step is to build the KL measurement and calibration pipelines, and benchmark on realistic mocks. We note that there are many existing methods in the WL cosmology toolbox that can be adapted for KiLeR. The baseline shape measurement tools for Stage-III surveys, e.g. \texttt{lensfit} \cite{Miller2013} and \texttt{ngmix} \cite{Sheldon2014}, are built on forward modeling. Extending forward modeling to slitless spectroscopy for kinematic reconstruction has been under active investigation over the past few years, especially in the JWST era \cite{2020A&A...633A..43O,2021ApJ...923..203S,2022zndo...7351572B,2024ApJ...976L..27N,2025MNRAS.543.3249D,2025arXiv251007369D,2025arXiv250315587S,2025ApJ...980...69L,2026MNRAS.546ag119D, 2026MNRAS.547ag437D,2026ApJ..1000..280A}.
By integrating the ground-based KL pipelines \cite{2020MNRAS.499.4591G, Pranjal2023,Pranjal2025,2026ApJ...999..222H} into the kinematic forward modeling, a Roman KL measurement pipeline can be built for pixel-to-shear inferences.
The shear calibration, however, can be more straightforwardly adapted from the existing responses-based methods, such as \texttt{metacal} \cite{Huff2017,Sheldon2017,Sheldon2020} and \texttt{anacal} \cite{2023MNRAS.521.4904L,LiAnaCal2025}. The slitless spectra are essentially 2D images with a color-dependent PSF. Therefore, by applying an artificial distortion to not only the images but also the spectra, we can characterize the KL shear response as we do in WL.
Another key infrastructure, the realistic mocks of Roman spectroscopy, is a shared requirement \citep[e.g.][]{2022ApJ...928....1W,2024arXiv241208883G} with galaxy clustering and peculiar velocity cosmology, further reducing the development cost of KiLeR.

Looking beyond Roman, there are abundant opportunities to expand the KiLeR sample and enhance its constraining power. Euclid \cite{Euclid} and the China Space Station Telescope \citep[CSST;][]{CSST} will deliver Roman-like WL+slitless spectroscopy surveys with a much larger area, but shallower depth and lower spectral resolution. In particular, the UV and optical coverage of CSST spectroscopy provides complementary low-redshift ($z<0.6$) coverage to Roman and Euclid, which is crucial for constraining dark energy evolution. Moreover, Multi-Object Spectrograph (MOS) surveys can expand the KL sample through follow-up spectroscopic observations \cite{Xu2024} to reduce kinematic reconstruction uncertainties. As the required source number density of KL is relatively low, the KL follow-up can serve as a spare-fiber program for high-fiber-density MOS, such as the Multiplexed Survey Telescope \citep[MUST;][]{MUST,MUST_LSS}, Spec-S5 \citep{Spec_s5}, and the Wide Field Spectroscopic Telescope \citep[WST;][]{WST}. The KL forecasts on DESI \cite{Xu2024} and WST \cite{Camera2026} have shown promising outcomes. Radio surveys also provide potential low-redshift complementary KL samples through HI kinematics \cite{2006ApJ...650L..21M,2015MNRAS.451..383W,Huang2025}.

In conclusion, we establish the Kinematic Lensing Ratio as a robust, precision probe of cosmic expansion history. A key lesson learned from the successful implementation of WL cosmology is that a strong science case can drive sustained development of technical capabilities and infrastructures, which eventually meet the uncertainty control requirements that initially seem challenging. The KiLeR method provides a unique science case for pursuing the maturation of KL measurements and advancing the understanding of galaxy kinematics, which will benefit not only dark energy cosmology but also a wide range of astrophysics and fundamental physics.

\begin{acknowledgments}
  This work was supported in part by the U.S. Department of Energy, Office of Science, Office of High Energy Physics, under award No. DESC0009959.
  We thank Kyle Dawson, Martin Kilbinger, Pengfei Li, Shu-Rui Lin, Wentao Luo, and Zheng Zheng for their encouragement and insightful discussions, especially Zheng Zheng for proposing the acronym ``KiLeR''. We also thank the authors of \cite{Xu2023} for releasing the redshift distribution of the Roman KL sample.
  We used AI in coding and language polishing. All codes and chains produced for this work will be made publicly available upon publication.
\end{acknowledgments}

\bibliographystyle{apsrev4-2}
\bibliography{main}% Produces the bibliography via BibTeX.

\appendix*

\section{End matter: Details of Forecast}
\subsection{Source sample characterization}
The number density and redshift distributions are motivated by applying the following KL selection criteria to the COSMOS Mock Catalog \citep{Jouvel2009}, which is constructed from deep HST imaging and spectroscopy and is designed for realistic forecasts of space-based surveys,
\begin{enumerate}
  \item \textbf{Resolved and high SNR emission lines}: at least one of the H$\alpha$, H$\beta$, or [OIII] lines that trace the kinematics of galaxy disks is resolved within the wavelength range of the Roman grism and has a 7$\sigma$ detection;
  \item \textbf{Resolved and high SNR imaging}: half-light radius $r_{1/2}>0.1"$ and z-band magnitude $m_z<24.5$.
\end{enumerate}
These selections yield $n_g = 8$ arcmin$^{-2}$, but we follow \cite{Xu2023} and use $4$ arcmin$^{-2}$ to be conservative, given that the uncertainty of the H$\alpha$ luminosity function can vary the estimate from $4$ to $12$ arcmin$^{-2}$ \cite{Zhai2019, Saito2020}. The KL sample is then dominated by star-forming disk galaxies. As the lensing effect is physically agnostic to galaxy intrinsic properties, the KL sample does not need to be representative of all galaxies for an unbiased cosmological inference.
The shape noise $\sigma_{\gamma}\simeq 0.035$ is estimated by running a forward-modeling measurement pipeline on the mocks of both spectra and images, as detailed in \cite{Pranjal2023} and the Appendix B of \cite{Xu2023} and confirmed by independent analysis in \cite{2020MNRAS.499.4591G,2021MNRAS.502.5612G,2026ApJ...999..222H}.
This shape noise budget includes the velocity measurement uncertainty, the intrinsic scatter of the TFR \citep[e.g.][]{DESI.EDR.TFsample,DESI.DR1.TFsample}, and the shape measurement uncertainties, and is dominated by the kinematic part \citep{2015MNRAS.451.2161D,2021MNRAS.502.5612G,2021ApJ...922..116D,2023ApJ...945...88D, 2026ApJ...999..222H}.

\subsection{Data vector}

\begin{figure}
  \centering
  \includegraphics[width=246pt]{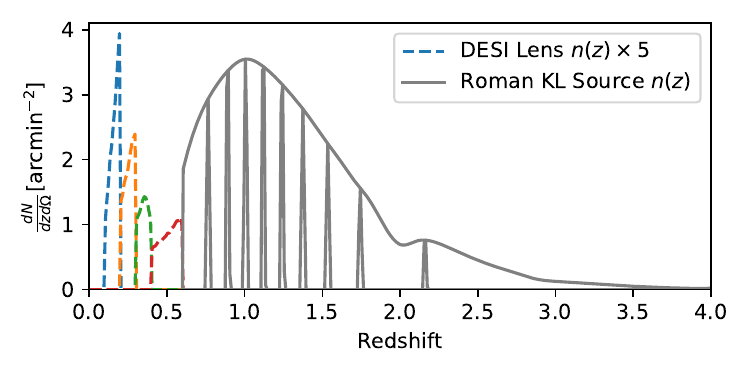}
  \caption{\label{fig:nz} The redshift distributions of the DESI-like lens samples (dashed lines) and Roman KL source sample (solid lines) used in the forecast. The lens samples include three DESI BGS samples at $z<0.4$ with different absolute magnitude cuts and one DESI LRG sample at $0.4<z<0.6$. The Roman KL source sample is divided into 10 tomographic bins from $z=0.6$ to $z=3$. }
\end{figure}

We use DESI-like spectroscopic samples at $z<0.6$ for lenses. As defined in \cite{2024MNRAS.533..589Y}, we adopt three different $r$-band absolute magnitude cuts ($M_r<-19, -20, -21$) for the DESI BGS sample at three redshift bins ($z\in[0.1,0.2], [0.2,0.3], [0.3,0.4]$), respectively. We note that this is not required for KiLeR, but the HOD parameters that are well studied in \cite{2024MNRAS.533..589Y} make it a convenient choice for our forecast.
For the DESI LRG sample at $z\in[0.4,0.6]$, we use the HOD parameters from \cite{2024MNRAS.530..947Y}. The redshift distributions of the lens and source samples are shown in Fig.~\ref{fig:nz}.

Our lens sample choice does not represent an optimal lens selection for KiLeR, but rather a conservative assumption for reliable signal and covariance estimation. In particular, we only include lenses at $z<0.6$ to avoid any ambiguity from overlapping lens and source redshift distributions. Although DESI itself does not have a large overlap area with Roman, it represents a typical target selection of MOS surveys such as 4MOST Cosmology Redshift Survey \citep{4MOST,2026MNRAS.545f2116V}, which can provide spectroscopic lens samples over the full Roman footprint.

As for the angular binning, the 1-arcmin small-scale cut is adopted to avoid the highly non-linear regime and ensure the validity of the analytical covariance estimation.

The data vectors are computed using the halo model framework in \texttt{AUM}. We assume the Tinker08 halo mass function \cite{Tinker08}, the Duffy08 mass-concentration relation \cite{Duffy08}, and NFW density profiles \cite{NFW} for both central and satellite galaxies. With galaxy-matter correlation function $\xi_{\rm gm}$ computed from the HOD models, we calculate the projected surface density profiles $\Sigma(R)$ for each lens sample:
\begin{equation}
  \Sigma(R) = \bar{\rho}_{\rm m} \int \left[1+\xi_{\rm gm}\left(\sqrt{R^2/(1+z)^2 + \chi^2}\right)\right] {\rm d}\chi,
\end{equation}
where $\bar{\rho}_{\rm m}$ is the mean matter density. We then convert $\Sigma(R)$ to the tangential shear profile via $\gamma_t(\theta) = \Delta\Sigma(R=\theta D_l) \langle\Sigma_{\rm crit}^{-1}\rangle$, where the geometric lensing efficiency is averaged over the redshift distributions using angular diameter distances:
\begin{equation}
  \langle\Sigma_{\rm crit}^{-1}\rangle = \frac{4\pi G}{c^2} \int {\rm d}z_l n_l(z_l) D_l \int {\rm d}z_s n_s(z_s) \frac{D_{ls}}{D_s}.
\end{equation}
The KiLeR observable is constructed by taking ratios of $\gamma_t$ between a source bin $s_i$ and the reference bin $s_1$ for each radial bin and lens bin separately. This results in a data vector with 15 radial bins $\times$ 4 lens bins $\times$ 9 source bin pairs = 540 elements.

\subsection{Covariance matrix}

The covariance matrix of the tangential shear profiles is computed analytically using \texttt{OneCovariance}, summing the Gaussian (G), connected non-Gaussian (NG), and super-sample covariance (SSC) terms calculated based on the halo model:
\begin{equation}
  \mathbf{C} = \mathbf{C}_{\rm G} + \mathbf{C}_{\rm NG} + \mathbf{C}_{\rm SSC}.
\end{equation}
We emphasize that the covariance budget of KiLeR differs fundamentally from standard weak lensing. Due to the high precision of the KL sample (suppressed shape noise), the shape noise term only dominates the diagonal elements at very small scales. The error budget at medium-to-large scales is instead dominated by the sampling variance (included in $\mathbf{C}_{\rm G}$). We refer readers to \cite{OneCovariance,2020MNRAS.497.2699F} for detailed derivations of each covariance term.

\subsection{Likelihood and sampling}

The theoretical prediction for the KiLeR observable, which enters the likelihood, depends purely on the ratio of lensing efficiencies and the multiplicative bias parameters $m$:
\begin{equation}
  R^{l}_{s_i,s_1} = \frac{\langle\Sigma_{\rm crit}^{-1}\rangle_{l,s_i}}{\langle\Sigma_{\rm crit}^{-1}\rangle_{l,s_1}} \frac{1+m_{s_i}}{1+m_{s_1}}.
\end{equation}
We employ a Gaussian likelihood for the posterior sampling. Although the ratio of two Gaussian variables is not strictly Gaussian, we follow \cite{Sun2023} to check a few representative cases and confirm that the Gaussian approximation is valid in our forecast sample.

\subsection{Effect of magnification}
\label{sec:magnification}
The magnification effect, or the difference between shear and reduced shear, is a systematic effect that we ignored in our baseline forecast.
Instead of shear, the reduced shear $g=\gamma/(1-\kappa)$ is the actual observable in weak lensing, where $\kappa$ is the lensing convergence $\kappa=\Sigma/\Sigma_{\rm crit}$. The reduced shear introduces a magnification bias to the KiLeR observable,
\begin{equation}
  R^{l}_{s_i,s_1} = \frac{\langle\Sigma_{\rm crit}^{-1}\rangle_{l,s_i}}{\langle\Sigma_{\rm crit}^{-1}\rangle_{l,s_1}} \frac{1 - \Sigma_{l}(\theta)\langle\Sigma_{\rm crit}^{-1}\rangle_{l,s_1}}{1 - \Sigma_{l}(\theta)\langle\Sigma_{\rm crit}^{-1}\rangle_{l,s_i}}.
\end{equation}
This effect reintroduces a dependence on the lens mass distribution, potentially biasing the cosmological inference. In our baseline forecast setup, if no correction or cut is applied, the magnification bias can cause $\sim 1.4\sigma$ bias on $w_0$ and $\sim 0.6\sigma$ on $w_a$ for KiLeR.

However, there are several ways to mitigate this effect. First, we can restrict the radial range to large scales where the magnification effect is not significant. By simply removing the three smallest-scale bins, where the difference between shear and reduced shear is $>0.5\%$, the bias on $w_0-w_a$ drops to an unbiased level ($\sim 0.2\sigma$), with a FoM loss of $\sim50\%$, reaching FoM of 16.5 (60) for KiLeR(+BAO).
An optimal scale cut can be determined from real data by examining the radial dependence of the shear ratios.
 Second, we can model the lens mass distribution using the measured shear profiles. Although this approach compromises the geometric nature of KiLeR, we note that the mass-modeling uncertainty here propagates only as a second-order effect in the ratio observables, and the impact is expected to be much smaller compared to that in the $3\times2$pts analysis. Instead of marginalizing over the mass modeling parameters as nuisance parameters, the hierarchical modeling can be a safer practice. We first fit and correct for magnification with mass profile templates motivated by simulations or halo models \citep[similar to][]{2024A&A...685A.167E}, then check the residual radial dependence and do the scale cut.
 Finally, we can measure the magnification $\kappa$ directly from e.g. the small-scale clustering \citep[e.g.][]{2024PhRvD.110j3538Q,2025PhRvD.112d3522Q}, and the size or flux changes \citep[e.g.][]{2010MNRAS.405.1025M, 2011PhRvD..84j3004V,2024ApJ...973..102X} of the background sources. A joint analysis combining magnification and shear ratios can also enhance the cosmological constraining power.
 As these are beyond the scope of this paper, we leave detailed investigations of the mitigation strategies to future work.

\end{document}